\documentclass[lettersize,journal]{IEEEtran}
\usepackage{amsmath,amsfonts}
\usepackage{algorithmic}
\usepackage{algorithm}
\usepackage{array}
\usepackage[caption=false,font=normalsize,labelfont=sf,textfont=sf]{subfig}
\usepackage{textcomp}
\usepackage{stfloats}
\usepackage{url}
\usepackage{verbatim}
\usepackage{graphicx}
\usepackage{cite}
\usepackage{booktabs}
\usepackage[export]{adjustbox}
\usepackage{booktabs}
\usepackage{multirow}
\usepackage{threeparttable}
\usepackage{threeparttablex}
\usepackage{longtable}
\usepackage{multirow}
\usepackage{graphicx}
\usepackage{orcidlink}
\begin{document}

\title{GeoTEAM: A Geospatial Tangible User Interface for Exploration and Visual Analysis of Migration Data}

\author{Karen Penaranda Valdivia \,\orcidlink{0000-0002-9883-0816},~\IEEEmembership{Graduate Student Member,~IEEE,\and \ Nujaimah Ahmed \,\orcidlink{0009-0004-2938-3890},\and \ Aswah Butt \, \orcidlink{0009-0008-2122-1757}, \and \ Nao Nagashima \, \orcidlink{0009-0004-7668-0417}, \and \ Sarah Hoyos-Hoyos \, \orcidlink{0000-0003-1150-3909}, \ Krupa Shah \, \orcidlink{0009-0001-3567-4174}, \and \ Elise van den Hoven \, \orcidlink{0000-0002-0888-1426}, ~\IEEEmembership{Member,~IEEE,  \and \ Robert McLeman \, \orcidlink{0000-0001-9593-1606}, \and \ Roozbeh Manshaei \, \orcidlink{0000-0001-9336-5831}, \and \ Gabby Resch \, \orcidlink{0000-0002-2842-111X}, \and \ Jamy Li \, \orcidlink{0000-0003-2440-9719}, ~\IEEEmembership{Member,~IEEE, \and and \ Ali Mazalek \, \orcidlink{0000-0003-0293-5435}}}}
    
\thanks{\textit{(Karen Penaranda Valdivia and Nujaimah Ahmed contributed equally to this work.)} \textit{(Corresponding author: Karen Penaranda Valdivia.)} This work involved human subjects in its research. Approval of all ethical and experimental procedures and protocols was granted by the Toronto Metropolitan University (TMU) Research Ethics Board under Application No. 2025-138 and performed in line with TCPS 2 (2022). 

Karen Penaranda Valdivia, Nujaimah Ahmed, Aswah Butt, Nao Nagashima, Krupa Shah, Dr. Roozbeh Manshaei, and Dr. Ali Mazalek are with the Synaesthetic Media Lab, Toronto Metropolitan University, Toronto, ON M5G 1P5, CA (e-mail: karen.penaranda@torontomu.ca).

Sarah Hoyos-Hoyos and Dr. Robert McLeman are with the Department of Geography and Environmental Studies, Wilfrid Laurier University, Waterloo, ON N2L 3C5, Canada.

Dr. Elise van den Hoven is with the School of Computer Science, University of Technology Sydney, Sydney, Ultimo NSW 2007, AU. 

Dr. Gabby Resch is with the Faculty of Business and Information Technology, Ontario Tech University, Oshawa, ON L1G 0C5, Canada.

Dr. Jamy Li is with the School of Computing, Engineering, \& The Built Environment, Edinburgh Napier University, Edinburgh EH11 4BN, UK. 

The authors were affiliated with the above institutions at the time the work was performed.

This work has been supported by the Social Sciences and Humanities Research Council, the Canada First Research Excellence Fund, the Canada Research Chairs program, the Canada Foundation for Innovation, and the Ontario Ministry of Research and Innovation.

}}

\markboth{\parbox{6.5in}{\raggedright THIS WORK HAS BEEN SUBMITTED TO THE IEEE FOR POSSIBLE PUBLICATION. COPYRIGHT MAY BE TRANSFERRED WITHOUT NOTICE, AFTER WHICH THIS VERSION MAY NO LONGER BE ACCESSIBLE.}}%
{Shell \MakeLowercase{\textit{et al.}}}

\maketitle

\begin{abstract}
Migration studies is an interdisciplinary field, requiring collaboration between researchers with varying levels of geospatial and quantitative data literacy. Geospatial tangible user interfaces (GTUIs) offer promising opportunities for embodied and collaborative spatial exploration of migration data. Few GTUIs, however, provide real-time visual feedback of data values to help users collaboratively identify trends. To address this gap, we present GeoTEAM, a novel tangible system for real-time, dynamic regional exploration, co-designed with migration and HCI researchers. GeoTEAM features active tangible dials with built-in touchscreens designed to control time and navigate map layers for net migration and various environmental sub-drivers in a multi-surface environment with tabletop and wall displays. We evaluated this system with nine pairs of researchers possessing multi-expertise in geospatial data literacy. Qualitative findings show that our system promotes collaboration, intuitive sense-making, and increased user confidence, enabled by physical interaction and real-time feedback from the active tangibles.
\end{abstract}


\section{Introduction}
\begin{figure}[h]
  \centering
  \includegraphics[width=0.9\linewidth]{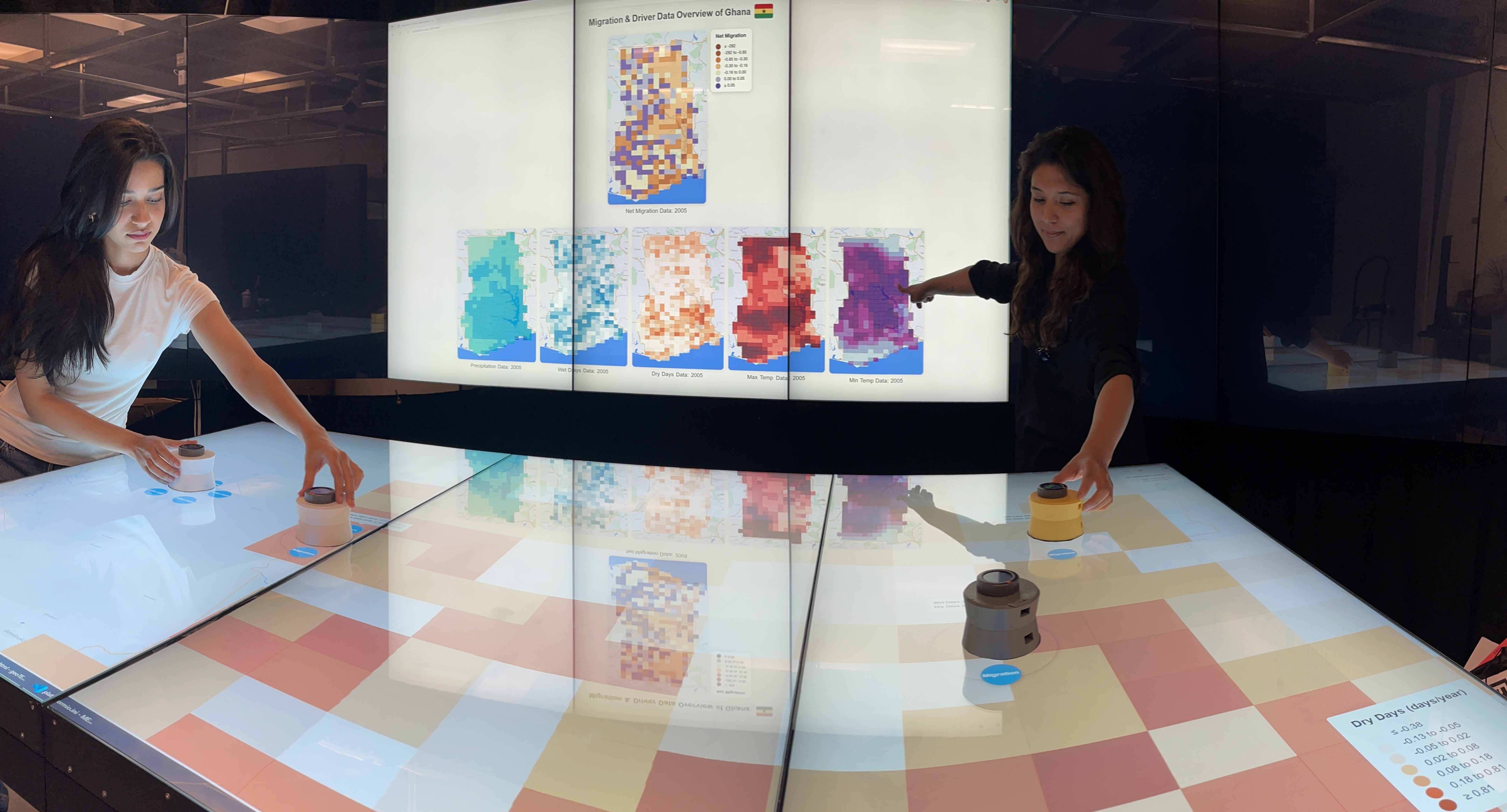}
  \caption{GeoTEAM's set-up consists of a (1) wall display, (2) tabletop display, (3) active tangibles, (4) legend, and (5) history of layers. The wall display hosts projections for net migration and associated sub-driver layers for a given time period. Users actively navigate map layers across the tabletop display with the tangibles, referring to the legend (bottom right corner), and the history of layers (top right corner).}
  \label{fig: set-up}
\end{figure}

\IEEEPARstart{R}{ecent} trends in international migration \cite{statistics_bundesamt_304_2025} have established the need for collaborative tools that enable interactive visualization and exploration of the complex and dynamic relationships driving human movement \cite{triandafyllidou_complex_2023}. Geospatial data science has moved beyond the scope of geographers and cartographers into interdisciplinary research concerns \cite{ziegler_need-finding_2023}, and migration researchers now leverage geospatial data analysis and visualization to explore migration drivers that include environmental (e.g., \cite{yang_utilizing_2017}) and socio-demographic factors (e.g., \cite{fatima_geospatial_2021, richards_gerrymandering_2014}). However, a recent study found that approximately 57\% of migration visualizations feature traditional line or bar charts that may have limited interactive and collaborative potential, and may not fully enable users to link the complex factors (i.e, environmental and socio-demographic contexts \cite{toronto_metropolitan_university_what_2024}) that influence migration with their visualized representations \cite{allen_conventions_2023}. Varying levels of geospatial modelling expertise can also lead to researchers working in isolation when using geographic information systems (GIS). Additionally, a common issue faced by GIS users is the lack of an expansive space to view mapped outputs \cite{ziegler_need-finding_2023}, and to perform individual or collaborative cartographic tasks (i.e., direct interactions, transformations, or analysis of spatial data   \cite{shakeri_hossein_abad_multi_2014}). This paper addresses these challenges by presenting GeoTEAM (refer to \IEEEpubidadjcol Fig. \ref{fig: set-up}), a collaborative \textbf{Geo}spatial \textbf{T}angible user interface for embodied \textbf{E}xploration and visual \textbf{A}nalysis of map layers for net \textbf{M}igration and its associated drivers. GeoTEAM uses active (i.e., non-passive, electronic \cite{brauer_active_2019}) tangible dials with built-in touchscreens, designed to navigate across a multi-surface environment. 

Prior work \cite{afkari_exploring_2020} highlighted how tangible tabletop interfaces could leverage collaboration by promoting teamwork, willingness to partake in group tasks, fairness in physical interactions, and shared attention on the tasks \cite{afkari_exploring_2020, ioannou_tabletops_2016}. Geospatial tangible user interfaces (GTUIs) have often been used for participatory urban planning, crisis management, and traffic simulation \cite{maquil_towards_2018}. However, GTUIs often include a large singular tabletop display (e.g., \cite{maquil_colortable_104, maquil_geospatial_2015, maquil_towards_2018, afkari_exploring_2020, anastasiou_design_2016, nagel_venice_2010}) and, despite the benefits of a large vertical wall display for sense-making \cite{andrews_space_2010}, collaboration \cite{jakobsen_up_2014}, and multi-modal gestures \cite{leon_eliciting_2024}, fewer works have implemented them (e.g., \cite{takahira_tangiblenet_2025, satriadi_tangible_2022}). At the time of writing, we found one previous work \cite{mahyar_ud_2016} resembling a GTUI that used both a tabletop and a wall display; however, unlike GeoTEAM, it did not use tangibles and relied on multi-touch interactions. 

Our co-design process, which included geography and migration experts, ensured that GeoTEAM has the potential to foster meaningful and innovative approaches toward data exploration in migration research — a domain poised to become significant for embodied and collaborative sense-making. This expands our contribution to the IEEE THMS community by promoting the use of tangible and embodied interfaces for real-world applications and opening up new possibilities for interdisciplinary migration research and design.

GTUI users have benefited from tangibles as major drivers of engaging and intuitive user-interface interactions \cite{angelini_move_2015,ens_uplift_2021,jones_twist_2015,valdes_exploring_2014}, which can facilitate handling of large geospatial datasets \cite{maquil_geospatial_2015} or support more distributed and embodied cognition \cite{hornecker_tei_2008}. However, a common limitation with these tangibles (e.g. wooden blocks \cite{maquil_geospatial_2015}, polyhedrons \cite{nagel_venice_2010}, tokens \cite{tateosian_tangeoms_2010}, and miniature globes \cite{satriadi_tangible_2022}) is that they may not provide visual feedback of menu selections and coordinate-based data values on the tangibles themselves. Provision of textual feedback in tangible user interfaces is important for enhancing usability, task performance, and learning \cite{anastasiou_questionnaire-based_2017}. 

GeoTEAM distinguishes itself from other GTUIs by meeting the gap for precise micro-regional exploration through the use of an MSE and active tangible dials that support real-time feedback of map-based values. GeoTEAM has four active tangible dials used to control parameters for time, net migration, and associated environmental sub-drivers. The tangible interactions in this study (i.e., interactions with the cartographic and system-perspective tasks \cite{shakeri_hossein_abad_multi_2014}) are somewhat related to a prior, currently unpublished elicitation study (refer to Table \ref{tab:Definitions}) with 36 participants possessing varying levels of geospatial modelling experience. The user tasks of this study are also collaboratively designed by a team of geography, HCI, visualization, and migration researchers. We implemented 2 scenarios for macro- and micro-regional exploration of map-based data, aiming to address these research questions: 
\begin{quote}
\textbf{RQ1:} How can a GTUI on a multi-surface environment (MSE) be designed to maximize usability and enhance user experience during interactive and collaborative exploration of migration data?
\\ \textbf{RQ2:} How do multiple users interact with and collaboratively explore migration using active tangibles on an MSE?
\end{quote}
To address RQ1 and RQ2, we recruited 18 participants with varying levels of self-described expertise in  geospatial data modelling; of these, 39\% (7 out of 18) worked in the field of migration studies. We conducted task-based study sessions using 2 scenarios—macro- and micro-regional exploration—with the 3 expert pairs, 3 mixed pairs, and 3 novice pairs. Qualitative content analysis revealed that participants across all levels of expertise found that the GeoTEAM system promotes collaboration, enhances sense-making with spatial information, and builds confidence through intuitive physical interactions and responsive real-time feedback from the active tangibles. 

\section{Related Work}
This section draws on past work covering multi-surface visualizations for large data, GTUIs, and active tangibles for large data exploration.

\subsection{Multi-Surface Visualizations for Large Datasets}

Multi-surface visualizations integrate multiple interactive displays, such as multi-touch tabletops and high-resolution wall-based screens, to provide an expansive visual space that supports collaborative exploration and sense-making of large datasets \cite{andrews_space_2010, leon_eliciting_2024,jakobsen_up_2014}. Liu et al. \cite{liu_effects_2014} found that multiple large displays, particularly tabletop or wall-sized configurations, offer substantial benefits for users engaged in complex tasks, including enhanced peripheral awareness, reduced cognitive load, and the ability to visualize and manipulate large data within a shared virtual workspace. Shared large displays foster closer physical proximity and increased communication among collaborators which, in turn, leads to improved shared understanding and group engagement \cite{jakobsen_up_2014,chen_use_2021}. Kim et. al \cite{kim_collaboration_2013} observed that large-scale multi-touch displays create workspaces that require users to physically move around the surface, prompting active interaction and communication as they navigate through the system. Additionally, their findings indicate that the design of an asynchronous system further encourages users to coordinate and collaborate, as they individually take turns to complete tasks rather than make changes to the system simultaneously. 

While previous studies have explored the use of single interactive display visualization systems for analyzing large datasets across various domains \cite{chen_use_2021}, very few systems have integrated multiple display types with active tangibles to support a more enhanced collaborative environment for users. This gap is significant given the complexity of migration datasets, which require collaborative environments where multiple users can interpret, compare, and share insights about migration data and associated underlying drivers. To address this gap, GeoTEAM explores how tabletop and wall displays can both integrate active tangibles to support a unified and interactive visual workspace. 

\subsection{Geospatial User Interfaces}

GTUIs are systems that employ physical objects to handle and interact with maps and geospatial data on tabletop displays \cite{maquil_towards_2018, maquil_geospatial_2015}. Maquil et al. \cite{maquil_towards_2018} found that users with varying geospatial expertise benefited from GTUIs as they enabled face-to-face interactions and discussions, and their tangibles were often perceived as playful and motivating, which facilitated users' organization and comprehension of spatial data. Additionally, for large datasets, tangibles can aid users with filtering data  \cite{ahlberg_dynamic_1992,spindler_tangible_2010} as well as changing the level of abstraction \cite{elmqvist_hierarchical_2010, spindler_tangible_2010}, which are both essential to encouraging understanding — especially with users of varying geospatial modelling expertise. 

Previous research highlighting the flexibility of GTUIs and TUIs in geospatial domains primarily used solid tangible blocks or spheres that do not have active components, such as electronic dials or rotary encoders. To illustrate, \textit{Mitigation in Urban areas: Solutions for Innovative Cities} (MUSIC) by Maquil et al. \cite{maquil_geospatial_2015} was a GTUI for urban planning projects that employed a tabletop screen with wooden tangible blocks. \textit{Venice Unfolding} by Heidmann, Condotta, and Duval \cite{nagel_venice_2010} was a GTUI for architectural planning that employed a tabletop screen with polyhedron tangibles. \textit{Smart City Logistics} by Guerlain, Cortina, and Renault \cite{guerlain_towards_2016} was a GTUI for commercial and energy resource planning that employed a tabletop screen and tangible tokens. \textit{TanGeoMS} by Tateosian et al. \cite{tateosian_tangeoms_2010} was a TUI for landform analysis that employed map projections onto sand and building block tangibles. Yuan et al. \cite{yuan_study_2018} also assessed industrial and environmental needs with a GTUI on a tabletop using Internet of Things (IoT) blocks that dynamically saved data. \textit{Tangible Globes} by Satriadi et al. \cite{satriadi_tangible_2022} was a TUI that used small globes and AR to investigate the tangible-virtual interplay of geospatial data exploration. Finally, \textit{TangibleNet} by Takahira et al. was a TUI for making links between map-based nodes on a wall display with projected spatial data and magnet tangibles.  

Despite this diversity, most existing works did not explore how tangible dials, which may differ from other interaction tools, might be employed for cartographic data exploration. GeoTEAM addresses this gap by employing active tangibles with embedded dials and interactive screen displays, which may enhance geospatial visualizations by enabling multi-modal feedback through rotary or push-button interactions (e.g., rotating and map presenting). These interactions intuitively translate MSE tasks and techniques to accomplish GIS system tasks \cite{shakeri_hossein_abad_multi_2014}, and could also strengthen spatial precision for positioning and adjusting maps upon navigation \cite{shakeri_hossein_abad_multi_2014, fiebrink_dynamic_2009}. 

\subsection{Active Tangibles for Large Data Exploration}

Active tangibles can facilitate the handling and understanding of geospatial data layers (i.e., map-based representations of geographic data \cite{adelfio_gisualization_2019}) through visual feedback. In \textit{The ColorTable} \cite{maquil_colortable_104}, tangibles (i.e., colour objects) offered immediate visual feedback of spatial positioning by highlighting the configuration area with the same colour as the object placed on it. In \textit{Venice Unfolding} \cite{nagel_venice_2010}, the polyhedron tangibles provided visual feedback of menu facets (i.e., architecture, date, typology, or keywords) after the user placed or tilted them on a specific edge. Once a specific edge was selected, the map views were instantly updated. In \textit{TanGeoMS} \cite{tateosian_tangeoms_2010}, after the users moved the building block tangibles or sand, the system provided immediate visual feedback of changes to the physical terrain through projections of vivid colour maps or overlays. In \textit{Orbitia} by Afkari, Maquil, and Anastasiou \cite{afkari_exploring_2020}, the \textit{drone} and the \textit{marker} tangibles provided immediate visual feedback by highlighting the scanned areas, and the \textit{drone} tangibles also showcased hidden items like minerals or obstacles on the tabletop. 

Despite the diversity of visual feedback provided by these works' tangibles, much of their feedback focused on marking regional areas where the user was navigating. A shared gap (e.g., \cite{maquil_colortable_104, nagel_venice_2010, tateosian_tangeoms_2010, afkari_exploring_2020}) was that the tangibles did not showcase dynamic textual feedback corresponding to the exact values of the data layers as users navigated the tabletop. GeoTEAM's tangible dials address this by dynamically showcasing  micro-regional values of the map layers, which may help reduce the cognitive burden of map and legend integration (i.e., matching the colours of the legend with the map's values) \cite{nussbaumer_knaflic_chapter_2020}. 

\section{System Overview and Technical Implementation}

\begin{figure}[htbp]
\centering
\includegraphics[width=0.9\linewidth]{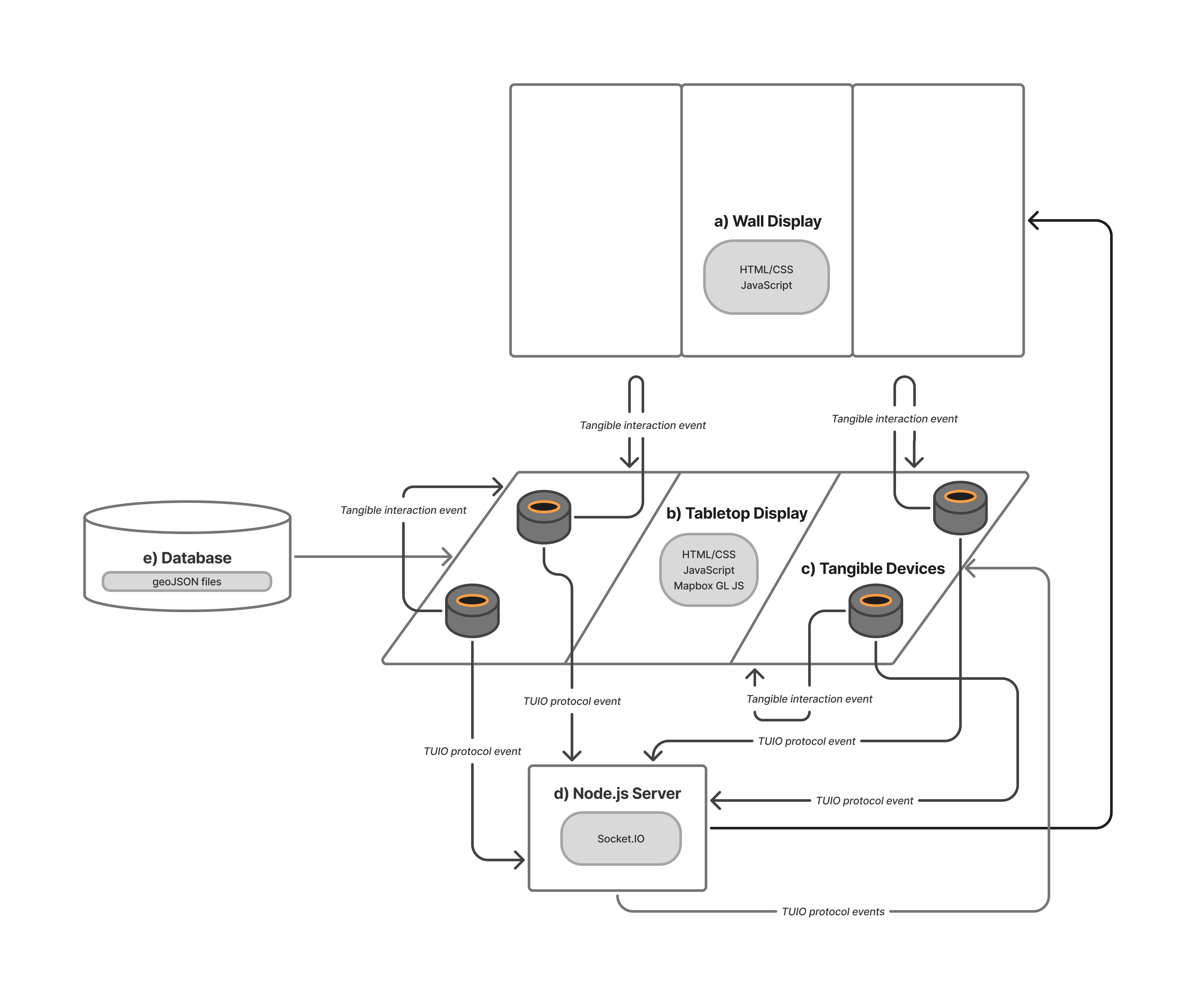}
\caption{GeoTEAM System Architecture consisting of a) wall display, b) tabletop display, c) tangibles, d) communication servers (Node.js server) and e) database that communicate with each other to process and present information.}
\label{system-architecture}

\end{figure}

GeoTEAM is explicitly designed to bridge physical action and digital data exploration via tangible dials, a tabletop display, and a wall display (refer to Fig. \ref{system-architecture}). GeoTEAM integrates object tracking and tangible recognition for multi-user, data-driven exploration and collaborative analysis across the MSE. The tangibles and their interactions were partially inspired by a prior, currently unpublished elicitation study carried out with 36 participants of varying self-described geospatial modelling expertise.

\subsection{Design of Tangibles and Interactions}

\begin{figure}[h]
  \centering
  \includegraphics[width=0.2\linewidth, height = 4 cm]{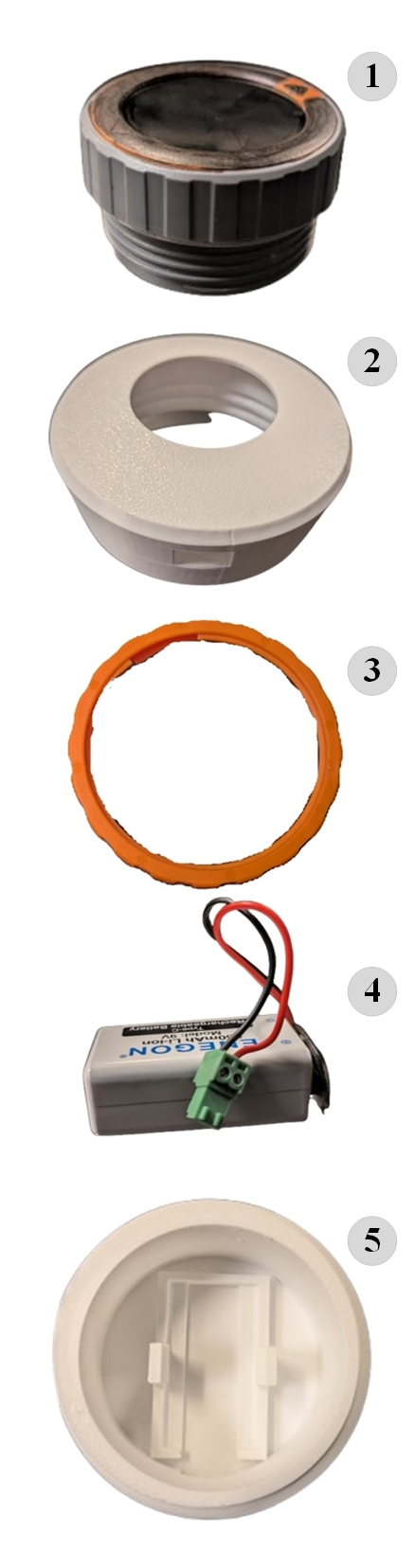}
  \caption{Exploded-view of the tangible dial. From top to bottom, the components are: (1) the M5Stack Dial, (2) the top-half of the cylindrical case, (3) the security ring of the M5Stack Dial, (4) 9V battery, and (5) the bottom-half of the cylindrical case with 9V battery holder.}
  \label{exploded-view}
\end{figure}

The tangibles are hand-held devices that include a M5Stack dial (refer to Fig. \ref{exploded-view}). The dials were selected for their programming flexibility and technical capabilities: ESP32-S3 chips programmable in Python and C++, Wi-Fi and Bluetooth connectivity for tabletop communication, and rotary encoders that enable rotating, zooming, and panning gestures. The embedded touchscreen present visual feedback for menu selections, map navigation, and map values.  

The tangible's cylindrical case was iteratively redesigned for ergonomics. Initially, the case had a pentagon shape, intended to represent the 5 macro-drivers of migration (i.e., environmental, socio-cultural, political, demographic, and economic \cite{toronto_metropolitan_university_what_2024}). However, few participants of the initial, exploratory currently unpublished elicitation study made this connection. Many expressed that a pepper-shaker-like shape would be more comfortable, which informed the current cylindrical case design.

The prior elicitation study's workflow roughly informed the design of the tangibles' interactions. In attempt to emulate a realistic routine of geospatial data exploration and analysis, this workflow was co-designed by researchers in migration, HCI, geography, and data science, and validated against prior literature  \cite{shakeri_hossein_abad_multi_2014} that mapped MSE tasks and techniques to GIS system tasks.  The unique gesture set (refer to Table \ref{tab:Definitions}) was coded by 3 researchers using GesturePTS \cite{li_gesturepts_2025}---a novel coding scheme we developed using pre-determined motion time-systems (i.e., MOST and MTM-1)---and the analysis was performed using the maximum-consensus and consensus-distinct ratio methodologies by Morris \cite{morris_reducing_2014}. Three gestures are pertinent to tangible dials: press for confirming menu selections, slide for viewing specific grid squares on the map during micro-regional exploration, and rotate to select specific years, menu options, or sub-driver layers.  
\subsection{Interaction Between the Tangibles and the MSE}

The tabletop display hosts a Mapbox-based webpage that shows the migration and environmental layers  (see b in Fig. \ref{system-architecture}) across different time periods. This webpage was developed using HTML, JavaScript, CSS, and Mapbox GL JS (a JavaScript library for developing interactive geospatial applications). The data for the layers is stored in GeoJSON files that are linked to a local database (see e in Fig. \ref{system-architecture}).  On the other hand, the wall display hosts snapshots of map overlays for migration and environmental drivers corresponding to time periods through a static web page implemented with HTML, JavaScript, and CSS (see a in Fig. \ref{system-architecture}). 

GeoTEAM offers 2 types of tangibles to interact with the tabletop display: (1) change time periods and (2) control the data layers and explore their map-based values in a scale of e.g. $25\,\mathrm{km}^2$. (Integration with the wall display is not in the scope of this paper.)  These interactions are visualized through custom pop circular menus that appear around the tangibles' periphery once placed on the tabletop. The precise location of the pop-up menus is facilitated by detection of the fiducial markers, pasted at the bottom of the tangibles.  To illustrate,  tabletop and wall displays  are made of 3 connected MultiTaction 55-Inch Ultra-Thin Bezel (MT553UTB) Cells, which have built-in infrared cameras equipped with a Hybrid Tracking Engine, and these employ the reacTIVision framework for fiducial marker-based object recognition via the Tangible User Interface Object (TUIO) protocol \cite{kaltenbrunner_reactivision_2007}.  

This real-time communication is coordinated by a Node.js server with a Socket.IO JavaScript library (refer to d in Fig. \ref{system-architecture}). The server's bridge module processes TUIO protocol events containing the tangible's coordinates and forwards them to the central Mapbox-based HTML web page on the tabletop display, which updates the tangible menu positions in real-time. To illustrate,  user interactions (i.e., rotation, touch input, or button press) trigger messages with the event details that update the tabletop display via a central Socket.IO websocket, changing the menu selections or activating and deactivating data layers based on the specific interaction. (The tangibles also provide auditory feedback of the confirmed selections, as the M5Stack dials have onboard buzzers that give short cues at various frequencies.) 

\begin{table*}
  \begin{threeparttable}
    \caption{Definitions and Illustrations Used to Code Participants' Proposed Gestures}
    \label{tab:Definitions}
    \begin{tabular}{p{2cm}p{9cm}p{5cm}}
      \toprule
      Gesture & Definition Used for Coding & Example Illustration\\
      \midrule
      Place & Changing position through the air without contacting a surface in the middle of the movement action. & \includegraphics[width = 2.5 cm, height = 0.7cm, valign=t]{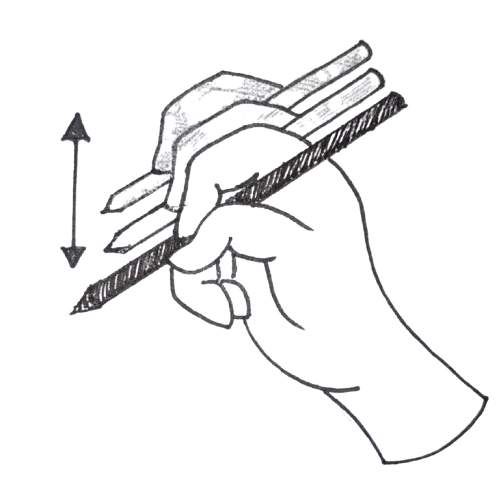}\\
      Press & Tap with the finger, touch with the finger, applying pressure with finger(s) onto object, except for swipe. & \includegraphics[height = 0.7cm, width = 5 cm, valign=t]{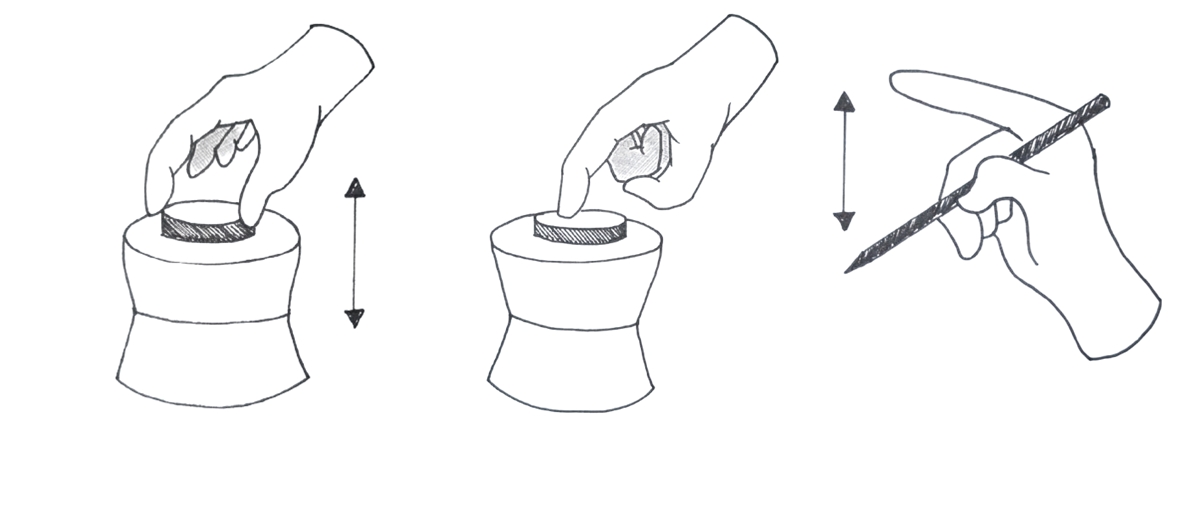}\\
      Slide & Tangibles keep surface contact, move the tangible(s) or stylus across the screen(s). & \includegraphics[width=5cm, height = 0.7cm, valign=t]{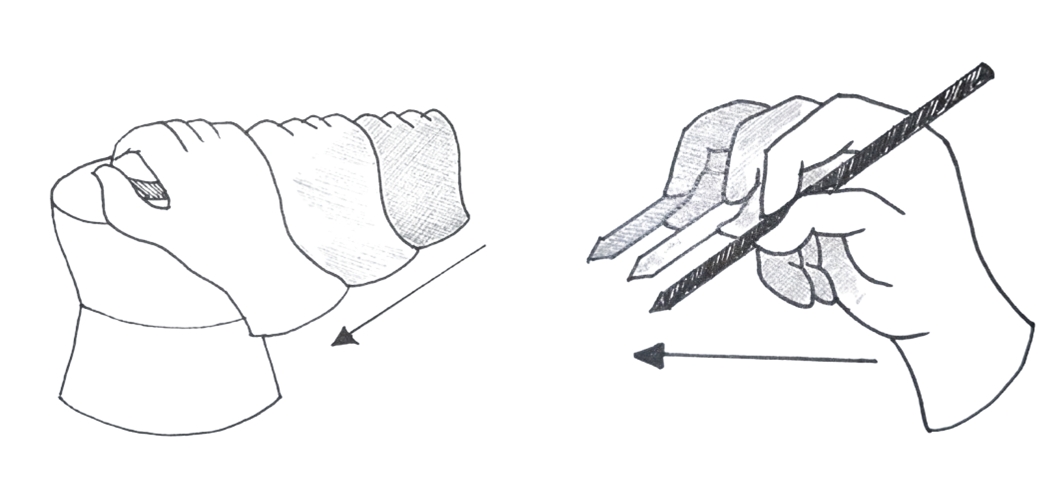}\\
      Rotate & Turn or pivot around a fixed point while holding a part (i.e., dial) or entire body of input devices (i.e., pentagon body or stylus). & \includegraphics[width=2.5cm, height = 0.7cm, valign=t]{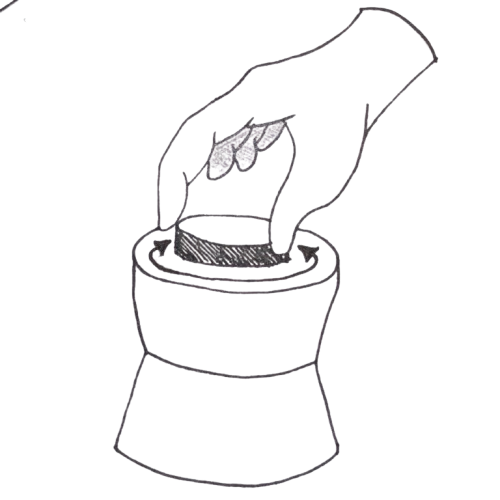}\\
      Lasso & Make a selection by creating a circular or loop-shaped path. The selection must be enclosed. Ensure surface contact with screen(s). & \includegraphics[width=2.5cm, height = 0.7cm, valign=t]{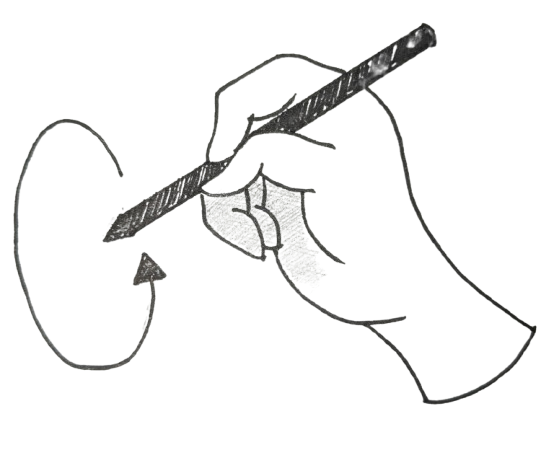}\\
      Draw & Make lines or shapes by moving across the screens. The selection does not have to be enclosed. & \includegraphics[width=2.5cm, height = 0.7cm, valign=t]{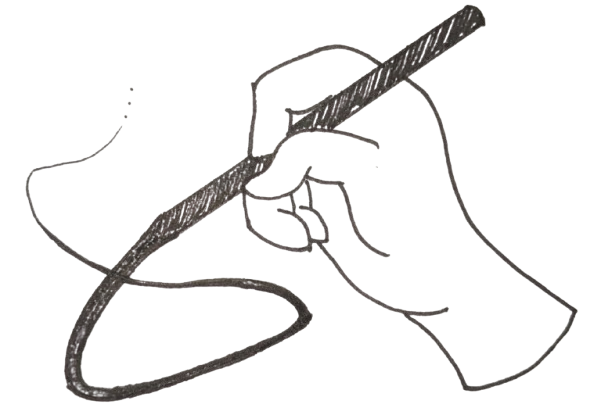}\\
      \bottomrule
    \end{tabular}
    \begin{tablenotes}[flushleft]
      \footnotesize
      \item[1] Tangibles are drawn as having cylindrical cases; in the currently unpublished elicitation study, they had pentagon cases.
      \item[2] Styluses were not used in this study; they were used in the currently unpublished elicitation study. 
    \end{tablenotes}
  \end{threeparttable}
\end{table*}

\section{Study Design}

We designed a task-based study to observe if, and how, users could interactively and collaboratively explore migration data at macro- and micro-regional scales using 2 scenarios (Table \ref{tab:scenarios}). 

\subsection{Participants}

We sought 18 participants with varying levels of self-described experience with geospatial data analysis and modelling to capture the diverse nature of migration research. Of the participants, 38.9\% (7/18) were recruited with the help of a university's migration research institute. The majority of them were 25 to 34 years old (61.1\% or 11/18), followed by 18 to 24 years old (38.9\% or 7/18). Completed education levels included bachelor's degrees (44.4\% or 8/18), followed by master's degrees (38.9\% or 7/18). Participants came from diverse careers, including migration and geography studies (44.4\%, 8/18), engineering (i.e., civil, architectural, geographic, industrial, and software) (38.9\% or 7/18), human-computer interaction and ergonomics (11.1\% or 2/18), and marketing (5.6\% or 1/18). Participants were randomly assigned into 9 pairs of novices, experts, and mixed groups based on expertise and shared time availability (refer to Table \ref{tab:Knowledge}). 

\subsection{Procedure}
Before the study began, participants confirmed their consent and authorized video- and audio- recording. Upon arrival, they were asked to complete a pre-task questionnaire to assess their experience with collaborative environments, interactive systems, and geospatial or migration data. These results were quantitatively analyzed to identify demographic trends. Then, participants were primed with a demonstration of GeoTEAM and its core functionalities.

There were 2 scenarios for exploration (refer to Table \ref{tab:scenarios}). In scenario 1 for macro-regional exploration, participants (A and/or B)  were given the task sequence of visually exploring environmental data layers for Ghana at the country-wide scale and across different time periods. In scenario 2 for micro-regional exploration, participants (A and/or B) were given the task  sequence of exploring environmental data layers for Ghana across different time periods from a pixel-by-pixel scale, where each pixel represented a $25\,\mathrm{km}^2$ grid square. The video and audio recordings of both scenarios were transcribed and, in combination with the logs of system interactions, were used to perform a thematic analysis.

After completing scenarios 1 and 2, each pair of participants was asked to provide open-ended feedback for the advantages and disadvantages of the prototype, alongside what features could be included in the next version. Lastly, participants were asked to complete a post-study questionnaire ranking collaboration, usefulness and sense-making, learnability and usability, tangible navigation, and tangible design. The questionnaire used a 7-point Likert scale, and was based on a combination of the System Usability Scale \cite{lewis_system_2018} and our own questions regarding specific features of the system. These results were statistically analyzed to detect any differences by expertise level. 

\subsection{Pre-Task Questionnaire Results}

In the pre-task questionnaire, out of the 9 expert and 9 novice participants, most experts (77.8\% or 7/9) reported working with map-based data regularly (i.e., on a daily, weekly, or monthly basis). Among the self-reported novices, 1 of 9 (11.1\%) reported frequently working with map-based data (weekly). All experts (100\% or 9/9) indicated that they work in collaborative environments either daily or weekly, compared to 7 of 9 novices (77.8\%). The majority of expert participants (77.8\% or 7/9) indicated that they are familiar with interactive systems, whereas almost half of the novice participants (44.4\% or 4/9) indicated the same. Only 1 novice reported frequent interaction with such systems, whereas the majority of expert participants had moderate experience with interactive systems. Expert and novice participants mentioned familiarity with a range of geospatial, design, and analysis software (e.g., Esri/ArcGIS, QGIS, AutoCAD, Civil 3D, Execution Analytics, Simply Analytics). These results provide context for interpreting participants’ performance and feedback.

\subsection{Thematic Results}

Two coders applied a hybrid approach, using inductive reasoning for participants' open-ended responses and deductive reasoning while analyzing gestures that may have been aligned with those from the prior elicitation study. Data familiarization involved: (1) transcribing and (2) reviewing videos of participants' interactions, (3) cross-referencing videos with gestures from the elicitation study when potentially useful, and (4) validating findings with logs from the study sessions. Inspired by Braun and Clarke's approach \cite{braun_using_2006}, after data-familiarization, the coders co-generated codes and identified 5 key themes related to user experience.  

\begin{table*}[ht]
    \centering
    \caption{Task Sequences for Scenarios 1 and 2}
    \label{tab:scenarios}
    \begin{tabular}{c p{8.3cm} p{8.3cm}}
        \hline
        \# & Scenario 1 Tasks & Scenario 2 Tasks \\
        \hline
        1 & A and B decide on a 5-year time interval to explore (i.e., 1985-1990 to 2005-2010). & A and B decide on a 5-year time interval to explore (i.e., 1985-1990 to 2005-2010). \\
        \hline
        2 & A (or B) uses the time tangible to select the chosen interval. & A (or B) uses the time tangible to select the chosen interval. \\
        \hline
        3 & B (or A) uses any of the 3 sub-driver tangibles to select the net migration layer. & B (or A) uses any of the 3 sub-driver tangibles to select a layer (e.g., precipitation, max. temperature, min. temperature, wet days, dry days). \\
        \hline
        4 & A and B identify any areas of importance or outliers visible in the visual data layer represented on the tabletop for the chosen sub-driver. & A and B place tangibles on specific map regions (pixels) to view the corresponding numerical data for that area. \\
        \hline
        5 & A and B take turns exploring the other environmental layers (i.e., precipitation, max. temperature, min. temperature, wet days, dry days) using the sub-driver tangibles. & A (or B) activates a new sub-driver layer that was not previously selected. \\
        \hline
        6 & A and B walk over to the wall display to review a summary of the results of all the layers previously explored on the tabletop display. & A and B place their tangibles back on the map regions to compare pixel-level data between 2 active layers. \\
        \hline
        7 & A and B discuss any patterns or connections between migration and environmental layers using the summary presented on the wall display. & NA \\
        \hline
        
    \end{tabular}
\end{table*}

\subsubsection{High-level Identification of Trends with Macro-Regional Exploration}

During macro-regional exploration, expert and novice participants effectively and consistently used GeoTEAM's colour-coded map layers across the MSE to  individually and collaboratively identify and compare trends. This was particularly useful to compare the net migration layer for a particular time period on the tabletop with all the sub-driver and net migration layers for another time period on the wall display. One expert (P2, G1) mentioned, ``[...] I mean that it is like the main key [...], the colours help to get the visualization part [of net migration]." Another novice (P7, G4) noted that being able to visualize an entire layer allowed for more immediate recognition of potential influences, noting this helped with, ``[...] Identifying trends, like looking at how one variable could impact another visually." The wall display facilitated collaborative pattern detection across seven out of nine groups (G1, G2, G3, G4, G5, G6, G7, and G8). In these groups, there were at least one or more instances where both participants identified areas of interest on the wall display based on colour changes and then lassoed or circled them with their fingers. 

\subsubsection{Confidence in Micro-Scale Exploration}

Micro-regional exploration facilitated participants' ability to identify spatial and temporal changes. Expert and novice participants tended to appreciate the ability to \textit{dial in} to precise locations at the pixel-level using the tangibles for more detailed analysis, which increased participants' confidence when interpreting connections between environmental and migration data.  One novice (P15, G8) noted, ``When you're working with data, especially and like trying to compare [time], like having the visual and kind of getting in there, that was nice." 

The micro-scale exploration also helped to overcome some challenges of stacking layers by mitigating visual distortion from merging layers. One novice participant (P15, G8) explained the advantage of micro-scale exploration as, ``[I like] Being able to see multiple things [written values] like migration, the wet days within the same year." Simultaneously displaying textual feedback of values for 2 or more data layers seemingly facilitated a clearer understanding of spatial relationships between different sub-drivers.

\subsubsection{Smooth Collaborative Dynamics}

The tangibles enabled turn-taking as pairs physically passed control to each other. Participants often highlighted how the passing of control minimized verbal interruptions and physical crowding. An expert (P12, G6) highlighted that GeoTEAM could be hosted at a  ``[local] Science Centre" to help \textit{kids} learn about group exploration in spatial contexts with tangibles. A novice participant (P17, G9)  also  highlighted that GeoTEAM could help level-set the learning process, saying, ``I thought it's like an interesting way to learn together. Like we learn how to use it together, and we learn how to recognize all the data together."  Many participants (P1 and P2, G1; P4, G2; P5, G3; P7, G4; P10, G5; P11, G6; P14, G7; P15, G8; and P17, G9) also mentioned that the large displays facilitated shared understanding of spatial trends, and seeing spatial data on a larger scale enabled their groups to have a broader view of patterns between migration and environmental layers. 

\subsubsection{Responsive Embodied Feedback}

Participants consistently valued the tangible’s immediate, real-time feedback, which reinforced a sense of direct physical control and streamlined their analytic workflow. For example, both novice (G8) and expert (P10, G5) participants noted that instant visual feedback reduced their cognitive load by minimizing reliance on legends and external data (e.g., it minimized the need to check a CSV-file for exact values). One participant (P17, G9) also mentioned that the tangibles increased collaborative transparency as unlike traditional mouses, participants could mutually observe each others' focus points. 

\subsubsection{Engaging and Intuitive Tangible Interactions}

One participant expressed that the tangible increased their engagement by creating an immersive and accessible user experience (P16, G9). GeoTEAM's collaborative design also fostered embodied interaction and open dialogue throughout the visual analysis tasks. In particular, the physical act of rotating the dial and touching the screen to make selections made data navigation intuitive and enjoyable, described as ``playing with the data" (P8, G4). Novice participants mentioned that, while there was an initial learning curve with the tangible interactions, they were able to quickly grasp how to use them after a few attempts. One expert participant (P1, G1) mentioned, ``...It's quite easy after using it once," along with a novice participant (P8, G4) noting, ``I think anybody who knows how to use a phone can use this device." The hands-on nature of the device helped lower the barrier for users unfamiliar with GIS, encouraging open-ended data exploration.

\subsection{Post-Study Questionnaire Results}

Participants completed a 30-question survey, answering each question with a 7-point Likert-scale (i.e., strongly disagree = 1, disagree = 2, somewhat disagree = 3, neither disagree nor agree = 4, somewhat agree = 5, agree = 6, strongly agree = 7).  Since we did not use the standard System Usability Scale, and instead used a hybrid approach with questions focused on understanding GTUI visualizations, this limited direct comparability with established usability rating benchmarks \cite{lewis_system_2018}. However, we used the Kruskal-Wallis test to compare the Likert-scale scores across 3 independent groups (experts, mixed pairs, and novices). Normality and homogeneity were not assumed. If a statistically significant difference among the experts, mixed pairs, and novices was found, Dunn's post-hoc analysis with Bonferroni correction was applied to determine which pairs of groups were statistically different. The 6 statistically significant differences identified with Dunn's post-hoc analysis with Bonferroni correction ($\alpha = 0.025$) include:
\begin{itemize}
    \item \textbf{Expert pairs vs novice pairs} for question 18 reading, ``I found the overall design of the tangible device visually appealing," wherein $p_{\text{expert~pairs vs novice~pairs}} = 0.0034 < 0.025$, $\mu_{\mathrm{expert~pairs}} =  6.67$, $\mu_{\mathrm{mixed~pairs}} =  5.67$, and $\mu_{\mathrm{novice~pairs}} = 4.67$. 
    \item \textbf{Expert pairs vs novice pairs} for question 19 reading, ``The system helped me to understand the dataset more clearly," wherein $p_{\text{expert~pairs vs novice~pairs}} = 0.0027 < 0.025$, $\mu_{\mathrm{expert~pairs}} = 6.67 $, $\mu_{\mathrm{mixed~pairs}} = 5.67 $, and $\mu_{\mathrm{novice~pairs}} = 4.67$.
    \item \textbf{Expert pairs vs novice pairs} for question 20 reading, ``The system supported my ability to identify patterns or relationships in the data," wherein $p_{\text{expert~pairs vs novice~pairs}} = 0.0085  < 0.025$, $\mu_{\mathrm{expert~pairs}} = 6.67$, $\mu_{\mathrm{mixed~pairs}} =  5.17$, and $\mu_{\mathrm{novice~pairs}} = 4.50$.
    \item \textbf{Expert pairs vs novice pairs} for question 22 reading, ``I found the system useful for completing the task at hand," wherein $p_{\text{expert~pairs vs novice~pairs}} = 0.0184  < 0.025$, $\mu_{\mathrm{expert~pairs}} = 6.50$, $\mu_{\mathrm{mixed~pairs}} =  5.17$, and $\mu_{\mathrm{novice~pairs}} = 4.67$.
    \item \textbf{Expert pairs vs mixed pairs} for question 24 reading, ``I would consider using a system like this in my own work or research," wherein $p_{\text{expert~pairs vs mixed~pairs}} = 0.0089  < 0.025$, $\mu_{\mathrm{expert~pairs}} = 6.50$, $\mu_{\mathrm{mixed~pairs}} =  4.17$, and $\mu_{\mathrm{novice~pairs}} = 5.17$.
    \item \textbf{Expert pairs vs mixed pairs} for question 27 reading, ``I found the visualizations of data on the tabletop useful for understanding the location of migration patterns,"  wherein $p_{\text{expert~pairs vs mixed~pairs}} = 0.0075  < 0.025$, $\mu_{\mathrm{expert~pairs}} = 6.50$, $\mu_{\mathrm{mixed~pairs}} =  5.17$, and $\mu_{\mathrm{novice~pairs}} = 6.00$.
\end{itemize}

\begin{table}
  \caption{Geospatial Data Modelling Knowledge by Group}
  \label{tab:Knowledge}
  \begin{tabular}{lll}
    \toprule
    Group \# & Participant \#  & Group Knowledge Level\\
    \midrule
    G1 & P1, P2& Expert\\ 
    G2 & P3, P4& Mixed (P3 is novice, P4 is expert) \\ 
    G3 & P5, P6& Mixed (P5 is expert, P6 is novice) \\
    G4 & P7, P8& Novice\\
    G5 & P9, P10 & Expert\\
    G6 & P11, P12 & Mixed (P11 is novice, P12 is expert)\\
    G7 & P13, P14& Expert \\
    G8 & P15, P16&  Novice\\
    G9 & P17, P18& Novice\\
  \bottomrule
\end{tabular}
\end{table}

\begin{figure}
    \centering
    \includegraphics[width=1\linewidth]{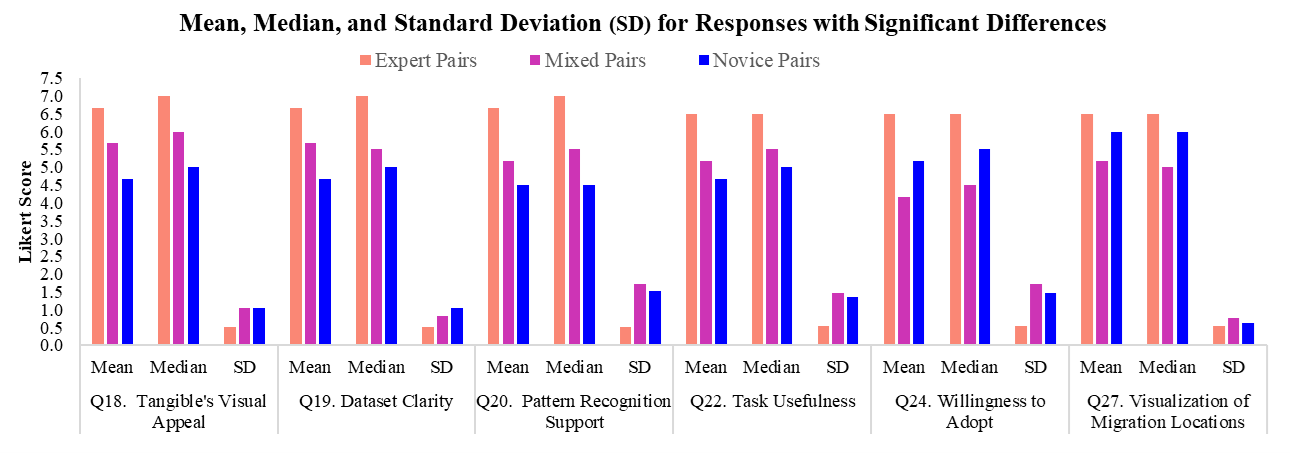}
    \caption{Mean, median, and standard deviation (SD) values for responses with significant differences based on expertise level.}
    \label{graph}
\end{figure}

The corresponding mean, median, and standard deviation values for these six significant differences based on the pairs' expertise levels are shown in Fig. \ref{graph}.

The statistically significant results indicate a difference in user experience based on expertise. In comparison to novice pairs, expert pairs consistently rated GeoTEAM as more visually appealing, helpful for analytical clarity, and effective for data comprehension and pattern detection. Mixed pairs' responses for visual appeal and analytical clarity tended to align with experts, but they diverged in terms of adaptability for person, work or research, and the helpfulness of the visualizations to identify the location of patterns. In these latter categories, novices tended to align more closely with experts. 

\section{Discussion}
This section covers design implications, including future work, and considers the limitations of this study.

\subsection{Design Implications}
This subsection examines areas for improvement and directions for future work, focusing on use-cases influenced by expertise, familiarity with tangible interactions, annotation needs, and integration of the tangibles with the wall display. 

\subsubsection{Use Cases for Macro- and Micro-Regional Exploration Varied Across Expertise Levels.}

We found that tangible interactions for navigating across different time periods or sub-driver layers were associated with expertise levels. This aligns with earlier findings that experts tend to apply systematic, fine-grained approaches to handling geospatial data \cite{brouwer_burg_it_2017}, whereas novices tend to apply \textit{bottom-up} approaches \cite{grueau_towards_2016}, characterized by a more gradual analysis process. For example, expert pairs (G1, G5, and G7) favoured micro-regional exploration, frequently using the tangibles to switch across multiple time periods for net migration or sub-driver layers. Some expert pairs (G1, G7) noted that macro-regional exploration was too \textit{high-level} and it did not offer enough insights for detecting system-wide trends. In contrast, some novice pairs (G4 and G9) and some novices in mixed pairs (P3, G2 and P6, G3) favoured macro-regional exploration to explore one layer (e.g., one net migration or one sub-driver layer) at a time. 

Nevertheless, GeoTEAM also showed promise for supporting novices to move beyond a \textit{bottom-up} approach by helping contextualize numerical values in a broader spatial framework. Three novices noted (P6, G3; P16, G8 and P18, G9) that the micro-regional exploration could be initially overwhelming without a previous understanding of how to interpret the data values, but the combination of macro-regional views and the wall-display projection facilitated visual connections across the data. This observation highlights how GeoTEAM fills a gap in the literature by demonstrating that a GTUI featuring an MSE with active tangibles that provide dynamic data feedback can facilitate geospatial understanding among multi-expertise users.

To enhance macro- and micro-regional exploration for novices and experts, the tangibles could provide additional quantitative and contextual information. First, to enhance the macro-regional exploration for experts, 2 experts (P2, G1 and P12, G6) proposed adding auto-correlation functions (e.g., Moran's I) as tangible menu options to help identify trends. On the other hand, 2 experts (P5, G3 and P12, G6) and 2 novices (P6, G3 and P16, G8) recommended adding pictorial or text information pop-ups as new tangible menu options. This information could be related to historical events, the influences of one sub-driver on another/others, or how to interpret legends. With regard to the latter, some novices (P7, G4; P11, G6; and P16, G8) and 1 expert (P12, G6) noted that legends may not be inherently intuitive (i.e., associating colours with values may be difficult), and it would be useful to add a ``help feature" for value clarification.

\subsubsection{Collaborative and Distributed Interactions}
We noted a  potential risk related to pairing experts with novices in collaborative systems is that the former may exhibit territoriality or frustration with the latter's questions \cite{thom-santelli_what_2010} — especially if experts do not see themselves as core users of the system. GeoTEAM aimed to address this by promoting mixed-focus collaboration, referring to individual and joint activities \cite{dewan_mixed-focus_2010}, via the use of multiple tangibles and a large-sized MSE. The tabletop and wall displays provided ample individual and shared workspaces, and the micro-regional view sustained engagement and sparked curiosity among mixed-expertise pairs. To illustrate, taking turns to make time or layer selections, followed by individual exploration with tangibles, enabled mixed pairs to interact and discuss findings. One novice (P11, G6) noted that it was helpful to ``do their own thing" and then show or discuss their findings with their partner. While individual behaviours may have influenced this outcome, all mixed pairs (G2, G3, and G6) reported benefits from having shared interactions with the tangibles.  

Future work will expand on collaborative and distributed interactions by enabling simultaneous selections of time and driver layers. We will investigate the use of split-screens and the possibility of having 2 modes for individual and shared explorations for individual and group autonomy.

\subsubsection{User Familiarization Required for Tangible Dial Interactions}

Expert, novice, and mixed pairs noted that there was a learning curve with using the tangible's dial when exploring different layers, primarily due to its button's unclear physical affordances and the need to follow a 2-stage process for choosing new layers. First, the button is hard to distinguish from the dial's outermost ring as it lies flat against it. In an attempt to make the button more obvious, we applied feedback from the elicitation study to add an arrow on top of the button and colour it differently from the rest of the ring. However, it still took most groups (G2-G9) 2 to 3 tries to locate the button and apply the right amount of force.  Second, the 2-stage process to change the layers involved: (1) tapping on the dial's screen to deactivate the current layer on display, and (2) pressing the button to return to the menu options. However, almost all participants (G1-G9) forgot to complete this routine on the first attempt, typically gaining confidence after the second try.

To enhance the usability of the dial's button, we will explore adding force feedback to complement visual feedback from the arrow on the tangible's ring. Greis et al. \cite{greis_extending_2019} note that users intuitively associate applied pressure with system certainty. Different tactile responses when pressing the ring versus the button may enhance users' spatial awareness and facilitate location differentiation between these components. 

To simplify layer de-selection, we will follow Nielsen's interface heuristics to minimize user effort and align with users' natural expectations \cite{nielsen_heuristic_1990}. Based on recommendations from some pairs of different expertise (G1, G3, G6, and G9), we will eliminate the use of the button for returning to the layer menu. Instead, we will attempt to automatically display the layer menu after the user has tapped on the dial's screen to deactivate the current layer. 

\subsubsection{Annotation and Documentation Needs}

At least once in the study, all participants (G1-G9) used finger gestures to circle, highlight, or point to areas of perceived importance. Similar finger motions are described by Shakeri et. al \cite{shakeri_hossein_abad_multi_2014} as useful for pattern detection and geospatial data interpretation. For macro-regional exploration, groups (G2, G5, and G8) circled entire cities to identify regional trends. For micro-regional exploration, groups (G1, G2, G6, and G8) circled pixels on the tabletop, and then searched for the corresponding pixels on the wall display to track period changes. To support communication and memory of these findings, one expert (P2, G1) suggested a pop-up annotation \textit{notepad} feature that could be activated from the tangible's menu. With this in mind, opportunities exist to include menu options for writing or free-hand selections (e.g., lassoing, circling, or drawing) that are activated with the tangible dials, as well as to integrate other annotation tools like styluses or finger-enabled multi-touch on the MSE. 

\subsubsection{Integration of Tangibles to the Wall Display to Visualize Changes Across Time}

In addition to annotation, some groups (G1, G2, G5, G8, and G9) expressed a desire to have a pictorial record of their chosen layers. While these groups often found the wall display projections advantageous for seeing all sub-driver layers and net migration for a particular time period, they wanted additional customization. Namely, they recommended that we complement the wall display with the addition of a dynamic time-series display, which would be created by enabling users to take \textit{snapshots} of multiple layers for different time periods through the tangibles. This aligns with  Shakeri et. al's \cite{shakeri_hossein_abad_multi_2014} recommendation that collaborative devices should allow users to share identified patterns without interrupting others' work. While one user posts something on the wall display, the other(s) could continue working on the tabletop. Future work will include integrating the tangibles with the wall display to enable screenshot and pasting features across displays. 

\subsection{Limitations}
At the time of the study, the tangibles' fiducial markers communicated solely with the tabletop, while the wall display was used to project net migration and all its associated sub-drivers for one particular time period. We will use the results of this study to refine our system and design new interactions for the tangibles and other input modalities, like styluses, for the wall display.

This study's results are based on insights from 18 participants with diverse geospatial modelling backgrounds, highlighting GeoTEAM's strong potential for interdisciplinary collaboration and embodied interaction. Future studies with a larger participant pool and broader set of use cases could expand insights into GeoTEAM's interdisciplinary applicability for collaborative migration research.

\section{Conclusion}

In this study, we introduced GeoTEAM, a multi-surface geospatial tangible user interface system designed to empower users with varying levels of geospatial modelling expertise to collaboratively and interactively explore migration data through active tangible dials. GeoTEAM fosters cross-expertise engagement and collaboration, both of which can be enhanced by integrating tangibles with a wall display, streamlining tangible interactions (e.g., deactivating layers), and providing tangible menu options to support trend identification. By leveraging users' understanding of geospatial data as they engage in macro- and micro-regional exploration, and by employing intuitive tangible interactions and embodied feedback, GeoTEAM facilitates deeper cross-expertise understanding of migration patterns. In doing so, it responds to the growing need for interactive tools that will enable collaborative exploration, visualization, and interpretation of the complex factors that drive contemporary human migration.

\bibliographystyle{IEEEtran}
\bibliography{references}

@inproceedings{li_gesturepts_2025,
	address = {Vienna Austria},
	title = {{gesturePTS}: {Using} {Predetermined} {Time} {Systems} ({PTS}) from {Human} {Factors} {Engineering} for coding gesture proposals from elicitation studies},
	language = {English},
	booktitle = {Proceedings of the 2025 {IEEE} {International} {Conference} on {Systems}, {Man}, and {Cybernetics} ({SMC})},
	publisher = {IEEE},
	author = {Li, Jamy and Penaranda, Karen},
	year = {2025},
}

@inproceedings{greis_extending_2019,
	address = {France},
	title = {Extending {Input} {Space} of {Tangible} {Dials} and {Sliders} for {Uncertain} {Input}},
	url = {https://hal.science/hal-02413724},
	doi = {10.1145/3294109.3300985},
	language = {English},
	urldate = {2025-08-07},
	booktitle = {Proceedings of the {Thirteenth} {International} {Conference} on {Tangible}, {Embedded}, and {Embodied} {Interaction}},
	publisher = {ACM},
	author = {Greis, Miriam and Kim, Hyunyoung and Korge, Andreas and Coutrix, Céline and Schmidt, Albrecht},
	month = mar,
	year = {2019},
	pages = {7},
}

@article{braun_using_2006,
	title = {Using thematic analysis in psychology},
	volume = {3},
	issn = {1478-0887, 1478-0895},
	url = {http://www.tandfonline.com/doi/abs/10.1191/1478088706qp063oa},
	doi = {10.1191/1478088706qp063oa},
	language = {en},
	number = {2},
	urldate = {2025-08-07},
	journal = {Qualitative Research in Psychology},
	author = {Braun, Virginia and Clarke, Victoria},
	month = jan,
	year = {2006},
	pages = {77--101},
}

@inproceedings{dewan_mixed-focus_2010,
	address = {Berlin Germany},
	title = {Mixed-focus collaboration without compromising individual or group work},
	isbn = {978-1-4503-0083-4},
	url = {https://dl.acm.org/doi/10.1145/1822018.1822054},
	doi = {10.1145/1822018.1822054},
	language = {en},
	urldate = {2025-08-07},
	booktitle = {Proceedings of the 2nd {ACM} {SIGCHI} symposium on {Engineering} interactive computing systems},
	publisher = {ACM},
	author = {Dewan, Prasun and Agarwal, Puneet and Shroff, Gautam and Hegde, Rajesh},
	month = jun,
	year = {2010},
	pages = {225--234},
}

@article{brouwer_burg_it_2017,
	title = {It must be right, {GIS} told me so! {Questioning} the infallibility of {GIS} as a methodological tool},
	volume = {84},
	issn = {03054403},
	url = {https://linkinghub.elsevier.com/retrieve/pii/S0305440317300730},
	doi = {10.1016/j.jas.2017.05.010},
	language = {en},
	urldate = {2025-08-07},
	journal = {Journal of Archaeological Science},
	author = {Brouwer Burg, Marieka},
	month = aug,
	year = {2017},
	pages = {115--120},
}

@inproceedings{adelfio_gisualization_2019,
	address = {Denmark},
	title = {{GISualization}: visualized integration of multiple types of data for knowledge co-production},
	volume = {119},
	shorttitle = {{GISualization}},
	url = {https://www.tandfonline.com/doi/full/10.1080/00167223.2019.1605301},
	doi = {10.1080/00167223.2019.1605301},
	language = {en},
	urldate = {2025-02-01},
	booktitle = {Geografisk {Tidsskrift}-{Danish} {Journal} of {Geography}},
	publisher = {Taylor \& Francis},
	author = {Adelfio, Marco and Kain, Jaan-Henrik and Stenberg, Jenny and Thuvander, Liane},
	month = jul,
	year = {2019},
	pages = {163--184},
}

@incollection{nussbaumer_knaflic_chapter_2020,
	address = {USA},
	edition = {1},
	title = {Chapter {Five} {Think} {Like} a {Designer}},
	volume = {1},
	isbn = {978-1-119-62149-2},
	url = {https://learning.oreilly.com/library/view/storytelling-with-data/9781119621492/c05.xhtml},
	language = {en},
	urldate = {2025-07-30},
	booktitle = {Storytelling with {Data}},
	publisher = {Wiley},
	author = {Nussbaumer Knaflic, Cole},
	year = {2020},
	pages = {137},
}

@article{lewis_system_2018,
	title = {The {System} {Usability} {Scale}: {Past}, {Present}, and {Future}},
	volume = {34},
	issn = {1044-7318, 1532-7590},
	shorttitle = {The {System} {Usability} {Scale}},
	url = {https://www.tandfonline.com/doi/full/10.1080/10447318.2018.1455307},
	doi = {10.1080/10447318.2018.1455307},
	language = {en},
	number = {7},
	urldate = {2025-07-27},
	journal = {International Journal of Human–Computer Interaction},
	author = {Lewis, James R.},
	month = jul,
	year = {2018},
	pages = {577--590},
}

@inproceedings{mahyar_ud_2016,
	address = {Niagara Falls Ontario Canada},
	title = {{UD} {Co}-{Spaces}: {A} {Table}-{Centred} {Multi}-{Display} {Environment} for {Public} {Engagement} in {Urban} {Design} {Charrettes}},
	copyright = {https://www.acm.org/publications/policies/copyright\_policy\#Background},
	shorttitle = {{UD} {Co}-{Spaces}},
	url = {https://dl.acm.org/doi/10.1145/2992154.2992163},
	doi = {10.1145/2992154.2992163},
	language = {en},
	urldate = {2025-07-24},
	booktitle = {Proceedings of the 2016 {ACM} {International} {Conference} on {Interactive} {Surfaces} and {Spaces}},
	publisher = {ACM},
	author = {Mahyar, Narges and Burke, Kelly J. and Xiang, Jialiang (Ernest) and Meng, Siyi (Cathy) and Booth, Kellogg S. and Girling, Cynthia L. and Kellett, Ronald W.},
	month = nov,
	year = {2016},
	pages = {109--118},
}

@inproceedings{fiebrink_dynamic_2009,
	address = {Boston MA USA},
	title = {Dynamic mapping of physical controls for tabletop groupware},
	copyright = {https://www.acm.org/publications/policies/copyright\_policy\#Background},
	url = {https://dl.acm.org/doi/10.1145/1518701.1518778},
	doi = {10.1145/1518701.1518778},
	language = {en},
	urldate = {2025-07-14},
	booktitle = {Proceedings of the {SIGCHI} {Conference} on {Human} {Factors} in {Computing} {Systems}},
	publisher = {ACM},
	author = {Fiebrink, Rebecca and Morris, Dan and Morris, Meredith Ringel},
	month = apr,
	year = {2009},
	pages = {471--480},
}

@inproceedings{kim_collaboration_2013,
	address = {San Antonio Texas USA},
	title = {Collaboration on a large-scale, multi-touch display: asynchronous interaction and multiple-input use},
	copyright = {https://www.acm.org/publications/policies/copyright\_policy\#Background},
	shorttitle = {Collaboration on a large-scale, multi-touch display},
	url = {https://dl.acm.org/doi/10.1145/2441955.2441997},
	doi = {10.1145/2441955.2441997},
	urldate = {2025-07-13},
	booktitle = {Proceedings of the 2013 conference on {Computer} supported cooperative work companion},
	publisher = {ACM},
	author = {Kim, Henna and Snow, Sara},
	month = feb,
	year = {2013},
	pages = {165--168},
}

@inproceedings{kaltenbrunner_reactivision_2007,
	address = {Baton Rouge Louisiana},
	title = {{reacTIVision}: a computer-vision framework for table-based tangible interaction},
	copyright = {https://www.acm.org/publications/policies/copyright\_policy\#Background},
	shorttitle = {{reacTIVision}},
	url = {https://dl.acm.org/doi/10.1145/1226969.1226983},
	doi = {10.1145/1226969.1226983},
	urldate = {2025-07-13},
	booktitle = {Proceedings of the 1st international conference on {Tangible} and embedded interaction},
	publisher = {ACM},
	author = {Kaltenbrunner, Martin and Bencina, Ross},
	month = feb,
	year = {2007},
	pages = {69--74},
}

@article{chen_use_2021,
	title = {On the {Use} of {Large} {Interactive} {Displays} to {Support} {Collaborative} {Engagement} and {Visual} {Exploratory} {Tasks}},
	volume = {21},
	issn = {1424-8220},
	url = {https://www.ncbi.nlm.nih.gov/pmc/articles/PMC8704489/},
	doi = {10.3390/s21248403},
	number = {24},
	urldate = {2025-07-13},
	journal = {Sensors (Basel, Switzerland)},
	author = {Chen, Lei and Liang, Hai-Ning and Wang, Jialin and Qu, Yuanying and Yue, Yong},
	month = dec,
	year = {2021},
	pmid = {34960495},
	pmcid = {PMC8704489},
	pages = {8403},
}

@article{jakobsen_up_2014,
	title = {Up close and personal: {Collaborative} work on a high-resolution multitouch wall display},
	volume = {21},
	copyright = {https://www.acm.org/publications/policies/copyright\_policy\#Background},
	issn = {1073-0516, 1557-7325},
	shorttitle = {Up close and personal},
	url = {https://dl.acm.org/doi/10.1145/2576099},
	doi = {10.1145/2576099},
	language = {en},
	number = {2},
	urldate = {2025-07-08},
	journal = {ACM Transactions on Computer-Human Interaction},
	author = {Jakobsen, Mikkel R. and HornbÆk, Kasper},
	month = feb,
	year = {2014},
	note = {Publisher: Association for Computing Machinery (ACM)},
	pages = {1--34},
}

@misc{statistics_bundesamt_304_2025,
	type = {Government},
	title = {304 million migrants worldwide},
	url = {https://www.destatis.de/EN/Themes/Countries-Regions/International-Statistics/Data-Topic/Population-Labour-Social-Issues/DemographyMigration/Migrants.html},
	language = {en},
	urldate = {2025-07-07},
	journal = {Federal Statistical Office},
	author = {Statistics Bundesamt},
	year = {2025},
}

@article{triandafyllidou_complex_2023,
	title = {Complex {Migration} {Flows} and {Multiple} {Drivers}: {What} {Do} {We} {Know}?},
	volume = {1},
	issn = {1929-9915},
	language = {en},
	number = {2023/05},
	journal = {The Working Paper Series},
	author = {Triandafyllidou, Anna and Ghio, Daniela and Veronis, Luisa and McLeman, Robert},
	year = {2023},
	pages = {4--5},
}

@article{allen_conventions_2023,
	title = {The conventions and politics of migration data visualizations},
	volume = {25},
	issn = {1461-4448},
	url = {https://doi.org/10.1177/14614448211019300},
	doi = {10.1177/14614448211019300},
	language = {EN},
	number = {6},
	urldate = {2025-07-07},
	journal = {New Media \& Society},
	author = {Allen, William L},
	month = jun,
	year = {2023},
	note = {Publisher: SAGE Publications},
	pages = {1313--1334},
}

@inproceedings{anastasiou_questionnaire-based_2017,
	address = {Limassol Cyprus},
	title = {A {Questionnaire}-based {Case} {Study} on {Feedback} by a {Tangible} {Interface}},
	isbn = {978-1-4503-4904-8},
	url = {https://dl.acm.org/doi/10.1145/3038535.3038540},
	doi = {10.1145/3038535.3038540},
	language = {en},
	urldate = {2025-07-06},
	booktitle = {Proceedings of the 2017 {ACM} {Workshop} on {Intelligent} {Interfaces} for {Ubiquitous} and {Smart} {Learning}},
	publisher = {ACM},
	author = {Anastasiou, Dimitra and Ras, Eric},
	month = mar,
	year = {2017},
	pages = {39--42},
}

@inproceedings{maquil_colortable_104,
	address = {Bonn Germany},
	title = {The \textit{{ColorTable}}: a design story},
	isbn = {978-1-60558-004-3},
	shorttitle = {The \textit{{ColorTable}}},
	url = {https://dl.acm.org/doi/10.1145/1347390.1347412},
	doi = {10.1145/1347390.1347412},
	language = {en},
	urldate = {2024-07-30},
	booktitle = {Proceedings of the 2nd international conference on {Tangible} and embedded interaction},
	publisher = {ACM},
	author = {Maquil, Valérie and Psik, Thomas and Wagner, Ina},
	year = {104},
	pages = {97--104},
}

@article{yang_utilizing_2017,
	title = {Utilizing {Cloud} {Computing} to address big geospatial data challenges},
	volume = {61},
	issn = {01989715},
	url = {https://linkinghub.elsevier.com/retrieve/pii/S0198971516303106},
	doi = {10.1016/j.compenvurbsys.2016.10.010},
	language = {en},
	urldate = {2025-07-05},
	journal = {Computers, Environment and Urban Systems},
	author = {Yang, Chaowei and Yu, Manzhu and Hu, Fei and Jiang, Yongyao and Li, Yun},
	month = jan,
	year = {2017},
	pages = {120--128},
}

@article{richards_gerrymandering_2014,
	title = {The {Gerrymandering} of {School} {Attendance} {Zones} and the {Segregation} of {Public} {Schools}: {A} {Geospatial} {Analysis}},
	volume = {51},
	issn = {0002-8312, 1935-1011},
	shorttitle = {The {Gerrymandering} of {School} {Attendance} {Zones} and the {Segregation} of {Public} {Schools}},
	url = {https://journals.sagepub.com/doi/10.3102/0002831214553652},
	doi = {10.3102/0002831214553652},
	language = {en},
	number = {6},
	urldate = {2025-07-05},
	journal = {American Educational Research Journal},
	author = {Richards, Meredith P.},
	month = dec,
	year = {2014},
	pages = {1119--1157},
}

@article{fatima_geospatial_2021,
	title = {Geospatial {Analysis} of {COVID}-19: {A} {Scoping} {Review}},
	volume = {18},
	copyright = {http://creativecommons.org/licenses/by/3.0/},
	issn = {1660-4601},
	shorttitle = {Geospatial {Analysis} of {COVID}-19},
	url = {https://www.mdpi.com/1660-4601/18/5/2336},
	doi = {10.3390/ijerph18052336},
	language = {en},
	number = {5},
	urldate = {2025-07-05},
	journal = {International Journal of Environmental Research and Public Health},
	author = {Fatima, Munazza and O’Keefe, Kara J. and Wei, Wenjia and Arshad, Sana and Gruebner, Oliver},
	month = jan,
	year = {2021},
	note = {Number: 5
Publisher: Multidisciplinary Digital Publishing Institute},
	pages = {2336},
}

@inproceedings{ziegler_need-finding_2023,
	address = {Hamburg Germany},
	title = {A {Need}-{Finding} {Study} with {Users} of {Geospatial} {Data}},
	isbn = {978-1-4503-9421-5},
	url = {https://dl.acm.org/doi/10.1145/3544548.3581370},
	doi = {10.1145/3544548.3581370},
	language = {en},
	urldate = {2025-07-05},
	booktitle = {Proceedings of the 2023 {CHI} {Conference} on {Human} {Factors} in {Computing} {Systems}},
	publisher = {ACM},
	author = {Ziegler, Parker and Chasins, Sarah E.},
	month = apr,
	year = {2023},
	pages = {1--16},
}

@article{ioannou_tabletops_2016,
	title = {Tabletops for {Peace}: {Technology} {Enhanced} {Peacemaking} in {School} {Contexts}},
	volume = {19},
	issn = {1176-3647},
	shorttitle = {Tabletops for {Peace}},
	url = {https://www.jstor.org/stable/jeductechsoci.19.2.164},
	number = {2},
	urldate = {2025-06-22},
	journal = {Journal of Educational Technology \& Society},
	author = {Ioannou, Andri and Antoniou, Chrystalla},
	year = {2016},
	note = {Publisher: International Forum of Educational Technology \& Society},
	pages = {164--176},
}

@inproceedings{afkari_exploring_2020,
	address = {Sienna, Italy},
	title = {Exploring {Opportunities} of {Tabletop} {Interfaces} for {Promoting} and {Analysing} {Collaboration}},
	language = {en},
	booktitle = {{ETIS} 2020},
	publisher = {CEUR Workshop Workshop Proceedings},
	author = {Afkari, Hoorieh and Maquil, Valrie and Anastasiou, Dimitra},
	year = {2020},
	pages = {1--6},
}

@inproceedings{takahira_tangiblenet_2025,
	address = {Japan},
	title = {{TangibleNet}: {Synchronous} {Network} {Data} {Storytelling} through {Tangible} {Interactions} in {Augmented} {Reality}},
	shorttitle = {{TangibleNet}},
	url = {http://arxiv.org/abs/2504.04710},
	doi = {10.1145/3706598.3714265},
	urldate = {2025-06-15},
	booktitle = {Proceedings of the 2025 {CHI} {Conference} on {Human} {Factors} in {Computing} {Systems}},
	publisher = {ACM},
	author = {Takahira, Kentaro and Kam-Kwai, Wong and Yang, Leni and Xu, Xian and Fujiwara, Takanori and Qu, Huamin},
	year = {2025},
	note = {arXiv:2504.04710 [cs]},
	pages = {1--18},
}

@inproceedings{anastasiou_design_2016,
	address = {Manchester United Kingdom},
	title = {Design {Implications} for a {User} {Study} on a {Tangible} {Tabletop}},
	isbn = {978-1-4503-4313-8},
	url = {https://dl.acm.org/doi/10.1145/2930674.2935982},
	doi = {10.1145/2930674.2935982},
	language = {en},
	urldate = {2025-06-18},
	booktitle = {Proceedings of the {The} 15th {International} {Conference} on {Interaction} {Design} and {Children}},
	publisher = {ACM},
	author = {Anastasiou, Dimitra and Maquil, Valérie and Ras, Eric and Fal, Mehmetcan},
	month = jun,
	year = {2016},
	pages = {499--505},
}

@inproceedings{satriadi_tangible_2022,
	address = {New Orleans LA USA},
	title = {Tangible {Globes} for {Data} {Visualisation} in {Augmented} {Reality}},
	isbn = {978-1-4503-9157-3},
	url = {https://dl.acm.org/doi/10.1145/3491102.3517715},
	doi = {10.1145/3491102.3517715},
	language = {en},
	urldate = {2025-06-15},
	booktitle = {{CHI} {Conference} on {Human} {Factors} in {Computing} {Systems}},
	publisher = {ACM},
	author = {Satriadi, Kadek Ananta and Smiley, Jim and Ens, Barrett and Cordeil, Maxime and Czauderna, Tobias and Lee, Benjamin and Yang, Ying and Dwyer, Tim and Jenny, Bernhard},
	month = apr,
	year = {2022},
	pages = {1--16},
}

@article{elmqvist_hierarchical_2010,
	title = {Hierarchical {Aggregation} for {Information} {Visualization}: {Overview}, {Techniques}, and {Design} {Guidelines}},
	volume = {16},
	issn = {1941-0506},
	shorttitle = {Hierarchical {Aggregation} for {Information} {Visualization}},
	url = {https://ieeexplore.ieee.org/document/5184827},
	doi = {10.1109/TVCG.2009.84},
	number = {3},
	urldate = {2025-06-15},
	journal = {IEEE Transactions on Visualization and Computer Graphics},
	author = {Elmqvist, Niklas and Fekete, Jean-Daniel},
	year = {2010},
	pages = {439--454},
}

@inproceedings{ahlberg_dynamic_1992,
	address = {Monterey, California, United States},
	title = {Dynamic queries for information exploration: an implementation and evaluation},
	isbn = {978-0-89791-513-7},
	shorttitle = {Dynamic queries for information exploration},
	url = {http://portal.acm.org/citation.cfm?doid=142750.143054},
	doi = {10.1145/142750.143054},
	language = {en},
	urldate = {2025-06-15},
	booktitle = {Proceedings of the {SIGCHI} conference on {Human} factors in computing systems  - {CHI} '92},
	publisher = {ACM Press},
	author = {Ahlberg, Christopher and Williamson, Christopher and Shneiderman, Ben},
	year = {1992},
	pages = {619--626},
}

@inproceedings{spindler_tangible_2010,
	address = {Saarbrücken Germany},
	title = {Tangible views for information visualization},
	isbn = {978-1-4503-0399-6},
	url = {https://dl.acm.org/doi/10.1145/1936652.1936684},
	doi = {10.1145/1936652.1936684},
	language = {en},
	urldate = {2025-06-13},
	booktitle = {{ACM} {International} {Conference} on {Interactive} {Tabletops} and {Surfaces}},
	publisher = {ACM},
	author = {Spindler, Martin and Tominski, Christian and Schumann, Heidrun and Dachselt, Raimund},
	year = {2010},
	pages = {157--166},
}

@article{guerlain_towards_2016,
	title = {Towards a {Collaborative} {Geographical} {Information} {System} to {Support} {Collective} {Decision} {Making} for {Urban} {Logistics} {Initiative}},
	volume = {12},
	copyright = {https://www.elsevier.com/tdm/userlicense/1.0/},
	issn = {23521465},
	url = {https://linkinghub.elsevier.com/retrieve/pii/S2352146516000181},
	doi = {10.1016/j.trpro.2016.02.017},
	language = {en},
	urldate = {2025-06-08},
	journal = {Transportation Research Procedia},
	author = {Guerlain, Cindy and Cortina, Stéphane and Renault, Samuel},
	year = {2016},
	pages = {634--643},
}

@inproceedings{nagel_venice_2010,
	address = {Reykjavik Iceland},
	title = {Venice unfolding: a tangible user interface for exploring faceted data in a geographical context},
	isbn = {978-1-60558-934-3},
	shorttitle = {Venice unfolding},
	url = {https://dl.acm.org/doi/10.1145/1868914.1869019},
	doi = {10.1145/1868914.1869019},
	language = {en},
	urldate = {2025-06-04},
	booktitle = {Proceedings of the 6th {Nordic} {Conference} on {Human}-{Computer} {Interaction}: {Extending} {Boundaries}},
	publisher = {ACM},
	author = {Nagel, Till and Heidmann, Frank and Condotta, Massimiliano and Duval, Erik},
	month = oct,
	year = {2010},
	pages = {743--746},
}

@inproceedings{nielsen_heuristic_1990,
	address = {New York, NY, USA},
	series = {{CHI} '90},
	title = {Heuristic evaluation of user interfaces},
	isbn = {978-0-201-50932-8},
	url = {https://dl.acm.org/doi/10.1145/97243.97281},
	doi = {10.1145/97243.97281},
	urldate = {2025-02-16},
	booktitle = {Proceedings of the {SIGCHI} {Conference} on {Human} {Factors} in {Computing} {Systems}},
	publisher = {Association for Computing Machinery},
	author = {Nielsen, Jakob and Molich, Rolf},
	month = mar,
	year = {1990},
	pages = {249--256},
}

@inproceedings{maquil_geospatial_2015,
	address = {Spain},
	title = {A geospatial tangible user interface to support stakeholder participation in urban planning},
	url = {https://ieeexplore.ieee.org/document/7512215},
	urldate = {2024-11-25},
	booktitle = {2015 1st {International} {Conference} on {Geographical} {Information} {Systems} {Theory}, {Applications} and {Management} ({GISTAM})},
	publisher = {IEEE},
	author = {Maquil, Valérie and De Sousa, Luís and Leopold, Ulrich and Tobias, Eric},
	month = apr,
	year = {2015},
	pages = {1--8},
}

@inproceedings{thom-santelli_what_2010,
	address = {New York, NY, USA},
	series = {{CHI} '10},
	title = {What do you know? experts, novices and territoriality in collaborative systems},
	isbn = {978-1-60558-929-9},
	shorttitle = {What do you know?},
	url = {https://dl.acm.org/doi/10.1145/1753326.1753578},
	doi = {10.1145/1753326.1753578},
	urldate = {2025-02-01},
	booktitle = {Proceedings of the {SIGCHI} {Conference} on {Human} {Factors} in {Computing} {Systems}},
	publisher = {Association for Computing Machinery},
	author = {Thom-Santelli, Jennifer and Cosley, Dan and Gay, Geri},
	month = apr,
	year = {2010},
	pages = {1685--1694},
}

@misc{toronto_metropolitan_university_what_2024,
	type = {Educational},
	title = {What makes {MEMO} {Unique}},
	shorttitle = {Background},
	url = {https://www.torontomu.ca/memo/overview/background/},
	language = {en},
	urldate = {2024-12-25},
	journal = {Toronto Metropolitan University (TMU)},
	author = {Toronto Metropolitan University},
	year = {2024},
}

@article{morris_reducing_2014,
	title = {Reducing legacy bias in gesture elicitation studies},
	volume = {21},
	issn = {1072-5520, 1558-3449},
	url = {https://dl.acm.org/doi/10.1145/2591689},
	doi = {10.1145/2591689},
	language = {en},
	number = {3},
	urldate = {2024-09-13},
	journal = {Interactions},
	author = {Morris, Meredith Ringel and Danielescu, Andreea and Drucker, Steven and Fisher, Danyel and Lee, Bongshin and Schraefel, M. C. and Wobbrock, Jacob O.},
	month = may,
	year = {2014},
	pages = {40--45},
}

@article{maquil_towards_2018,
	title = {Towards a framework for geospatial tangible user interfaces in collaborative urban planning},
	volume = {20},
	issn = {1435-5949},
	url = {https://doi.org/10.1007/s10109-018-0265-6},
	doi = {10.1007/s10109-018-0265-6},
	language = {en},
	number = {2},
	urldate = {2024-09-09},
	journal = {Journal of Geographical Systems},
	author = {Maquil, Valérie and Leopold, Ulrich and De Sousa, Luís Moreira and Schwartz, Lou and Tobias, Eric},
	month = apr,
	year = {2018},
	pages = {185--206},
}

@inproceedings{andrews_space_2010,
	address = {New York, NY, USA},
	series = {{CHI} '10},
	title = {Space to think: large high-resolution displays for sensemaking},
	isbn = {978-1-60558-929-9},
	shorttitle = {Space to think},
	url = {https://dl.acm.org/doi/10.1145/1753326.1753336},
	doi = {10.1145/1753326.1753336},
	urldate = {2024-09-04},
	booktitle = {Proceedings of the {SIGCHI} {Conference} on {Human} {Factors} in {Computing} {Systems}},
	publisher = {Association for Computing Machinery},
	author = {Andrews, Christopher and Endert, Alex and North, Chris},
	month = apr,
	year = {2010},
	pages = {55--64},
}

@inproceedings{liu_effects_2014,
	address = {Toronto Ontario Canada},
	title = {Effects of display size and navigation type on a classification task},
	isbn = {978-1-4503-2473-1},
	url = {https://dl.acm.org/doi/10.1145/2556288.2557020},
	doi = {10.1145/2556288.2557020},
	language = {en},
	urldate = {2024-07-30},
	booktitle = {Proceedings of the {SIGCHI} {Conference} on {Human} {Factors} in {Computing} {Systems}},
	publisher = {ACM},
	author = {Liu, Can and Chapuis, Olivier and Beaudouin-Lafon, Michel and Lecolinet, Eric and Mackay, Wendy E.},
	month = apr,
	year = {2014},
	pages = {4147--4156},
}

@article{hornecker_tei_2008,
	title = {{TEI} goes on: {Tangible} and embedded interaction},
	volume = {7},
	number = {2},
	journal = {IEEE Pervasive Computing},
	author = {Hornecker, Eva and Jacob, Robert JK and Hummels, Caroline CM and Ullmer, Brygg and Schmidt, Albrecht and van den Hoven, Elise AWH and Mazalek, Ali},
	year = {2008},
	pages = {91--96},
}

@article{tateosian_tangeoms_2010,
	title = {{TanGeoMS}: {Tangible} {Geospatial} {Modeling} {System}},
	volume = {16},
	copyright = {https://ieeexplore.ieee.org/Xplorehelp/downloads/license-information/IEEE.html},
	issn = {1077-2626},
	shorttitle = {{TanGeoMS}},
	url = {http://ieeexplore.ieee.org/document/5613503/},
	doi = {10.1109/TVCG.2010.202},
	number = {6},
	urldate = {2024-07-30},
	journal = {IEEE Transactions on Visualization and Computer Graphics},
	author = {Tateosian, L and Mitasova, H and Harmon, B and Fogleman, B and Weaver, K and Harmon, R},
	month = nov,
	year = {2010},
	pages = {1605--1612},
}

@inproceedings{yuan_study_2018,
	address = {Jeju, Korea (South)},
	title = {Study of {Design} {Method} for {Tangible} {User} {Interface} in {IoT} {Paradigm}},
	isbn = {978-1-5386-5457-6},
	url = {https://ieeexplore.ieee.org/document/8650374/},
	doi = {10.1109/TENCON.2018.8650374},
	urldate = {2024-07-30},
	booktitle = {{TENCON} 2018 - 2018 {IEEE} {Region} 10 {Conference}},
	publisher = {IEEE},
	author = {Yuan, Zihong and Ng, Wee Siong and Tat Goh, Shen and Zhou, Yimin},
	month = oct,
	year = {2018},
	pages = {2071--2074},
}

@inproceedings{jones_twist_2015,
	address = {Barcelona, Spain},
	title = {Twist, {Shift}, or {Stack}? - {Usability} {Analysis} of {Geospatial} {Interactions} on a {Tangible} {Tabletop}},
	isbn = {978-989-758-099-4},
	shorttitle = {Twist, {Shift}, or {Stack}?},
	url = {http://www.scitepress.org/DigitalLibrary/Link.aspx?doi=10.5220/0005377601700177},
	doi = {10.5220/0005377601700177},
	urldate = {2024-07-30},
	booktitle = {Proceedings of the 1st {International} {Conference} on {Geographical} {Information} {Systems} {Theory}, {Applications} and {Management}},
	publisher = {SCITEPRESS - Science and and Technology Publications},
	author = {Jones, Catherine Emma and Maquil, Valérie},
	year = {2015},
	pages = {170--177},
}

@incollection{grueau_towards_2016,
	address = {Cham},
	title = {Towards {Geospatial} {Tangible} {User} {Interfaces}: {An} {Observational} {User} {Study} {Exploring} {Geospatial} {Interactions} of the {Novice}},
	volume = {582},
	isbn = {978-3-319-29588-6 978-3-319-29589-3},
	shorttitle = {Towards {Geospatial} {Tangible} {User} {Interfaces}},
	url = {http://link.springer.com/10.1007/978-3-319-29589-3_7},
	urldate = {2024-07-30},
	booktitle = {Geographical {Information} {Systems} {Theory}, {Applications} and {Management}},
	publisher = {Springer International Publishing},
	author = {Jones, Catherine Emma and Maquil, Valérie},
	editor = {Grueau, Cédric and Gustavo Rocha, Jorge},
	year = {2016},
	doi = {10.1007/978-3-319-29589-3_7},
	note = {Series Title: Communications in Computer and Information Science},
	pages = {104--123},
}

@article{ens_uplift_2021,
	title = {Uplift: {A} {Tangible} and {Immersive} {Tabletop} {System} for {Casual} {Collaborative} {Visual} {Analytics}},
	volume = {27},
	issn = {1941-0506},
	shorttitle = {Uplift},
	url = {https://ieeexplore-ieee-org.ezproxy.lib.torontomu.ca/document/9229116},
	doi = {10.1109/TVCG.2020.3030334},
	number = {2},
	urldate = {2024-07-25},
	journal = {IEEE Transactions on Visualization and Computer Graphics},
	author = {Ens, Barrett and Goodwin, Sarah and Prouzeau, Arnaud and Anderson, Fraser and Wang, Florence Y. and Gratzl, Samuel and Lucarelli, Zac and Moyle, Brendan and Smiley, Jim and Dwyer, Tim},
	month = feb,
	year = {2021},
	note = {Conference Name: IEEE Transactions on Visualization and Computer Graphics},
	pages = {1193--1203},
}

@inproceedings{shakeri_hossein_abad_multi_2014,
	address = {Dresden Germany},
	title = {Multi {Surface} {Interactions} with {Geospatial} {Data}: {A} {Systematic} {Review}},
	isbn = {978-1-4503-2587-5},
	shorttitle = {Multi {Surface} {Interactions} with {Geospatial} {Data}},
	url = {https://dl.acm.org/doi/10.1145/2669485.2669505},
	doi = {10.1145/2669485.2669505},
	language = {en},
	urldate = {2024-06-18},
	booktitle = {Proceedings of the {Ninth} {ACM} {International} {Conference} on {Interactive} {Tabletops} and {Surfaces}},
	publisher = {ACM},
	author = {Shakeri Hossein Abad, Zahra and Anslow, Craig and Maurer, Frank},
	month = nov,
	year = {2014},
	pages = {69--78},
}

@inproceedings{brauer_active_2019,
	address = {New York, NY, USA},
	series = {{MuC} '19},
	title = {An {Active} {Tangible} {Device} for {Multitouch}-{Display} {Interaction}},
	isbn = {978-1-4503-7198-8},
	url = {https://dl.acm.org/doi/10.1145/3340764.3344436},
	doi = {10.1145/3340764.3344436},
	urldate = {2024-06-12},
	booktitle = {Proceedings of {Mensch} und {Computer} 2019},
	publisher = {Association for Computing Machinery},
	author = {Brauer, Christoph and Ariza, Oscar and Steinicke, Frank},
	month = sep,
	year = {2019},
	pages = {439--444},
}

@article{angelini_move_2015,
	title = {Move, {Hold} and {Touch}: {A} {Framework} for {Tangible} {Gesture} {Interactive} {Systems}},
	volume = {3},
	copyright = {Copyright MDPI AG 2015},
	shorttitle = {Move, {Hold} and {Touch}},
	url = {https://www.proquest.com/docview/1721922778/abstract/AF613C93954D46A0PQ/1},
	doi = {10.3390/machines3030173},
	language = {English},
	number = {3},
	urldate = {2024-06-10},
	journal = {Machines},
	author = {Angelini, Leonardo and Lalanne, Denis and Hoven, Elise Vanden and Khaled, Omar Abou and Mugellini, Elena},
	year = {2015},
	note = {Num Pages: 173-207
Place: Basel, Switzerland
Publisher: MDPI AG},
	pages = {173--207},
}

@article{leon_eliciting_2024,
	title = {Eliciting {Multimodal} and {Collaborative} {Interactions} for {Data} {Exploration} on {Large} {Vertical} {Displays}},
	volume = {30},
	copyright = {https://creativecommons.org/licenses/by/4.0/legalcode},
	issn = {1077-2626, 1941-0506, 2160-9306},
	url = {https://ieeexplore.ieee.org/document/10309866/},
	doi = {10.1109/TVCG.2023.3323150},
	number = {2},
	urldate = {2024-06-10},
	journal = {IEEE Transactions on Visualization and Computer Graphics},
	author = {León, Gabriela Molina and Isenberg, Petra and Breiter, Andreas},
	month = feb,
	year = {2024},
	pages = {1624--1637},
}

@inproceedings{valdes_exploring_2014,
	address = {Toronto Ontario Canada},
	title = {Exploring the design space of gestural interaction with active tokens through user-defined gestures},
	isbn = {978-1-4503-2473-1},
	url = {https://dl.acm.org/doi/10.1145/2556288.2557373},
	doi = {10.1145/2556288.2557373},
	language = {en},
	urldate = {2024-06-10},
	booktitle = {Proceedings of the {SIGCHI} {Conference} on {Human} {Factors} in {Computing} {Systems}},
	publisher = {ACM},
	author = {Valdes, Consuelo and Eastman, Diana and Grote, Casey and Thatte, Shantanu and Shaer, Orit and Mazalek, Ali and Ullmer, Brygg and Konkel, Miriam K.},
	month = apr,
	year = {2014},
	pages = {4107--4116},
}

\vfill

\end{document}